\documentclass[10pt, conference, letterpaper]{IEEEtran}
\IEEEoverridecommandlockouts

\usepackage{cite}
\usepackage{amsmath,amssymb,amsfonts}
\usepackage{algorithmic}
\usepackage{graphicx}
\usepackage{textcomp}
\usepackage{xcolor}
\usepackage{array}
\usepackage{multirow}
\usepackage{tabularx}
\usepackage{stfloats}
\usepackage{algorithm2e}

\usepackage{listings}
\usepackage{xcolor}
\usepackage{subfigure}
\usepackage{comment}
\usepackage{multirow}
\usepackage{makecell}
\usepackage{url}
\usepackage{wrapfig}
\usepackage{caption} 
\usepackage{hyperref}

\definecolor{codegreen}{rgb}{0,0.6,0}
\definecolor{codegray}{rgb}{0.5,0.5,0.5}
\definecolor{codepurple}{rgb}{0.58,0,0.82}
\definecolor{backcolour}{rgb}{0.95,0.95,0.92}

\lstdefinestyle{mystyle}{
    backgroundcolor=\color{backcolour},   
    commentstyle=\color{codegreen},
    keywordstyle=\color{magenta},
    numberstyle=\tiny\color{codegray},
    stringstyle=\color{codepurple},
    basicstyle=\ttfamily\footnotesize,
    breakatwhitespace=false,         
    breaklines=true,                 
    captionpos=b,                    
    keepspaces=true,                 
    numbers=left,                    
    numbersep=5pt,                  
    showspaces=false,                
    showstringspaces=false,
    showtabs=false,                  
    tabsize=2
}

\def\BibTeX{{\rm B\kern-.05em{\sc i\kern-.025em b}\kern-.08em
    T\kern-.1667em\lower.7ex\hbox{E}\kern-.125emX}}
\begin{document}

\title{Toward Non-Expert Customized Congestion Control: Large Language Model-Assisted CCA Code Generation with eBPF Deployment%
\thanks{This paper appears in the Proceedings of the IEEE International
Conference on Communications (ICC) 2025 under the title
``Toward Non-Expert Customized Congestion Control.'' The peer-reviewed
ICC 2025 version was submitted on 8 November 2024. DOI:
\url{https://doi.org/10.1109/ICC52391.2025.11160790}.
\textcopyright~2025 IEEE. Personal use of this material is permitted.
Permission from IEEE must be obtained for all other uses, in any current
or future media, including reprinting/republishing this material for
advertising or promotional purposes, creating new collective works, for
resale or redistribution to servers or lists, or reuse of any copyrighted
component of this work in other works.}
}

\author{
\IEEEauthorblockN{Mingrui Zhang, Hamid Bagheri, Lisong Xu}
  \IEEEauthorblockA{
  School of Computing,
  University of Nebraska-Lincoln
  }
Email: mzhang23@huskers.unl.edu,\{bagheri, xu\}@cse.unl.edu
}

\maketitle

\begin{abstract}

General-purpose congestion control algorithms (CCAs) are designed to achieve general congestion control goals, but they may not meet the specific requirements of certain users.
Customized CCAs can meet certain users' specific requirements; however, non-expert users often lack the expertise to implement them.
In this paper, we present an exploratory non-expert customized CCA framework, named NECC, which enables non-expert users to easily model, implement, and deploy their customized CCAs by leveraging Large Language Models and the Berkeley Packet Filter (BPF) interface. 
To the best of our knowledge, we are the first to address the customized CCA implementation problem.
Our evaluations using real-world CCAs show that the performance of NECC is very promising, and we discuss the insights that we find and possible future research directions.

\end{abstract}


\section{Introduction}\label{sec:introduction}

Many congestion control algorithms (CCAs), such as Reno, Cubic, and BBR, have been proposed and deployed on the Internet. 
These CCAs are designed by network experts to achieve general congestion control goals, such as fairly allocating bandwidth among competing flows, maximizing network utilization, and avoiding network congestion.

These general-purpose CCAs, however, may not meet the specific requirements of certain users. For example, consider the following scenario: a live-streaming user on platforms like YouTube Live, Twitch, or Instagram Live. If the user wants to guarantee 2K-resolution streaming from home, where the bottleneck is the home Internet connection (i.e., the last-mile link), the user may need a customized CCA that allocates a sufficient amount of bandwidth to the 2K-resolution streaming flow, with the remaining bandwidth allocated among other flows. This is a customized CCA tailored for this specific user, prioritizing the streaming needs over the general fairness goal.

Live streaming users may have basic video knowledge (e.g., Standard, High Definition, 2K, and 4K); however, they may not possess other necessary expertise to model, implement, and deploy the customized CCAs. In this paper, we refer to these users as non-expert users, who may lack the following types of expertise: 
1)  Non-expert users may not have the expertise to model their streaming requirements into a customized CCA design. 
2) They may not have the expertise to implement the customized CCA design into deployable CCA source code. This process necessitates comprehensive domain knowledge in networking and operating systems to effectively translate ideas into functional source code. 3) They may not have the expertise to deploy the customized CCA implementation on their streaming computers. The CCA deployment usually involves modifying kernel configurations and source code files, compiling, and installing the new kernel — all of which require expertise in kernel programming.

In this paper, we present a non-expert customized CCA framework called NECC, which enables non-expert users to easily model, implement, and deploy their customized CCAs by leveraging Large Language Models (LLMs). While the proposed framework is exploratory, it is the first step toward automatic customized CCAs for non-expert users, and it shows very promising performance. To the best of our knowledge, our work is first to address this problem. The primary contributions of this paper are as follows: 

\textit{1)} We propose implementing customized CCAs using the code refinement method instead of directly generating the code from scratch. Essentially, we treat customized requirements as code refinement criteria, view the existing CCA code as needing refinement to fully meet these requirements, and then use an LLM to generate refined CCA code that aligns with the customized requirements. 

\textit{2)} We propose several network domain-specific techniques to address the potentially erroneous outputs of an LLM, such as additional network safety requirements for safe deployment on the Internet, network chain-of-thought (CoT) \cite{NEURIPS2022_9d560961} prompts to guide the LLM in refining the appropriate CCA functions and adjusting the correct CCA variables using correct units, and network feedback to iteratively refine the LLM outputs based on CCA performance. 

\textit{3)} We evaluate our proposed framework and these network domain-specific techniques for live streaming users using real-world CCAs, such as Linux Cubic, Reno, and others. The experimental results are very promising. Additionally, we discuss the insights that we have found and possible future research directions.

\section{Related Work}

\textbf{\textit{Automated CCA Design:}} Learning-based CCA studies use machine learning (ML) algorithms to generate optimal solutions, such as policy-switching strategies \cite{sage2023} and dynamic variable control \cite{jay2019deep}. Different from those works that focus on rule-level learning, we study the generation of system-level executable code. \cite{agarwal_towards} takes a step forward by synthesizing objectives and proposing a provably effective heuristic CCA design generation approach, which is subsequently implemented using a third-party library. Unlike this work, our approach conceals the CCA design stage from non-expert users and directly implements the customized requirements as Linux-compatible CCA code.

\textbf{\textit{LLM in Networking:}} LLMs have been recently employed in network management \cite{Network-Management-llm}, network configuration generation \cite{NetConfEval}, etc. While existing networking code generation techniques have primarily been utilized for bug fixing and enhancing management performance, our focus is on generating executable CCA code that is compatible with the Linux kernel. Recent studies \cite{The-Age-of-Generative-AI,zhou2024large} have highlighted the potential of using LLMs or Generative AI for various networking tasks, such as AI-assisted data collection, congestion prediction, and network design. Our work represents the first exploratory step in utilizing LLMs to assist with customized CCA code, which we believe is a promising avenue for future research.

\textbf{\textit{Code Generation and LLM:}} Recent code generation research has turned to evaluating LLMs for code generation \cite{2024duevaluating,CoderEval}. Iterative feedback approaches have been proposed to improve LLM output quality in \cite{NEURIPS2023_1b44b878, NEURIPS2023_7cc1005e, NEURIPS2023_91edff07}. These iterative methods typically address general mathematical reasoning and basic programming tasks. We focus on network-specific coding challenges and propose using emulation-based feedback in iterative revision. \cite{jimenez2024swebench} utilizes issue statements and codebases as input to evaluate pull request patch generation. Our work uses similar input but for a different purpose: we regard customized requirements as issues to generate fulfilling source code. \cite{agentless} generates multiple patches for locally specific buggy files. Our work generates multiple code candidates to address the probabilistic output of LLMs.

\section{Problem statement}\label{sec:problem-statement}

In this paper, we study how to design a framework to enable non-expert users to easily model, implement, and deploy their customized CCAs. As an exploratory framework, we consider only live-streaming users in this paper, which is a popular type of Internet users who have special requirements not met by current general-purpose CCAs. 

The framework interacts with a user to collect the specific requirements and home network information. For example, a user may request support for streaming a game with 2K resolution in a home with an Internet connection speed of 60 Mbps. The framework outputs a script that automatically installs customized CCA code on a computer.

The framework should achieve the following design goals: 
\begin{itemize}
    \item \textbf{Goal 1: Non-expert Users}.
 The users of the framework should not need substantial expertise in modeling, implementing, and deploying customized CCAs.  
 
 \item \textbf{Goal 2: Functional CCA Code}. 
 The framework should generate functional customized CCA code according to user requirements. 
 
 \item \textbf{Goal 3: Safe Network Deployment}. 
 The generated CCA code should meet user requirements while being safe for network deployment (e.g., without starving other flows in the home network and leading to congestion collapse).  

\end{itemize}

\section{Proposed Framework}\label{sec:proposed-solution}

\begin{figure}[tb]
    \centering
    \includegraphics[width=\linewidth]{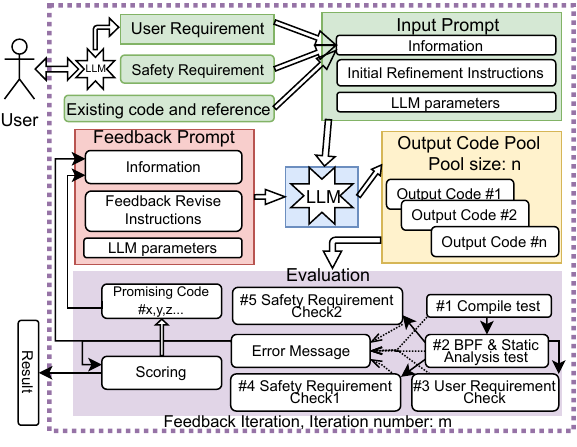}
    \caption{Framework of Non-Expert Customized CCA}
    \label{fig:System-Overview}
\end{figure}

In this section, we describe our proposed Non-Expert Customized CCA framework, called NECC, illustrated in Fig.~\ref{fig:System-Overview}. 

\subsection{Framework Overview}

NECC uses two LLMs. 1) A modeling LLM (at the top of the figure) chats with a user in natural language to collect the specific user requirements and home network information, and 2) An implementation LLM (in the middle of the figure) automatically generates customized CCA programs.

The input to the implementation LLM is a prompt, which consists of the following information: LLM parameter settings, a customized CCA design including both specific user requirements and additional requirements for safe network deployment, existing CCA code and references, and instructions to refine the existing code to meet all the requirements. 

The output of the implementation LLM is a pool of candidate CCA programs. Each candidate CCA program is evaluated against all the requirements. If at least one candidate CCA program meets all the requirements, NECC outputs such a program. Otherwise, NECC iteratively asks the implementation LLM to refine these candidate CCA programs based on the evaluation feedback.

\subsection{Design Choices for Achieving Design Goals}

NECC achieves the first design goal (i.e., non-expert users) by leveraging LLMs in modeling, implementing, and deploying customized CCA code. Specifically, the modeling LLM assists in collecting and modeling the specific user requirements, and the implementation LLM automatically generates customized CCA code using the BPF interface. The NECC framework also provides the user with an executable script to deploy the generated CCA code. NECC adopts the BPF interface because the BPF-based method deploys the user space CCA code by attaching to the kernel, which is transparent to the streaming applications. It also significantly simplifies the compilation and deployment process compared to traditional kernel-space methods.

The challenge in achieving the second design goal (i.e., functional CCA code) is that the implementation LLM may generate probabilistic and erroneous outputs. A probabilistic output means that a generated CCA program may be different at different times for the same prompt. An erroneous output means that a generated CCA program does not work for various reasons. NECC uses multiple techniques to address this challenge and generate functional CCA code:

\textit{1)} To lower the difficulty of code generation, NECC uses the code refinement method to modify existing CCA code instead of generating the code from scratch. This design choice is explained in detail in Section~\ref{sec:tcp-cca-generation-method}. 

\textit{2)} To address the potentially erroneous output of the implementation LLM, NECC uses network domain-specific CoT prompts to guide the implementation LLM in refining the correct CCA functions and adjusting the correct variables with the correct units. The CoT prompting is explained in detail in Section~\ref{sec:prompt}. 

\textit{3)} To mitigate the impact of the probabilistic output of the implementation LLM, NECC asks the LLM to generate a pool of candidate CCA programs instead of only one CCA program. 

\textit{4)} To mitigate the impact of the potentially erroneous output of the implementation LLM, NECC iteratively asks the LLM to refine CCA code based on the network domain-specific feedback. The feedback is explained in detail in Section~\ref{sec:feedback-iterative-revision}.

To achieve the third design goal (i.e., safe network deployment), NECC adds additional network safety requirements to the customized CCA design. These additional requirements are explained in detail in Section~\ref{sec:prompt}.

\section{Refinement-Based CCA Generation}\label{sec:tcp-cca-generation-method}

We choose the code refinement method to generate customized CCA code for the following reasons: 1) Recent code generation studies \cite{2024duevaluating, CoderEval} show that direct code generation using LLMs or traditional code generation models performs worse in long, complex class-level coding tasks. 2) There is already a rich body of CCA programs. For example, the Linux kernel contains the code of more than 15 different CCAs. 

We propose to change the customized CCA code generation problem into a CCA code refinement problem. Specifically, we treat customized requirements as a natural language description of code refinement criteria, view the existing CCA code as needing refinement to fully meet these requirements, and then
use an LLM to generate refined CCA code that aligns with the customized requirement.

NECC uses an LLM to conduct code refinement. This is because LLMs are ready to use and recent studies \cite{jiang2023impact} \cite{xia_automated_2023} show that they have at least the same level of performance as traditional code generation and code repair tools.

\section{Prompt of Implementation LLM}
\label{sec:prompt}

A prompt to the implementation LLM contains three types of network domain-specific information: instruction, CCA design information, and additional information.

\subsection{Instructions}

The instructions ask the implementation LLM to refine the existing CCA code to meet all the requirements in the customized CCA design. Our work is inspired by the CoT prompting proposed in recent code generation works~\cite{self-plan-cot,cot-in-code-gen} that focus on translating coding intent into step-by-step functional implementation descriptions. However, because we do not have CCA refinement examples to teach an LLM, our CoT prompts contain only domain-specific guidelines.

Our CoT prompts contain the following domain-specific guidelines to mitigate the potentially erroneous outputs of the implementation LLM: 1) A generated CCA program should compile. 2) The generated CCA program can be attached to the BPF interface. 3) The LLM should refine the appropriate CCA functions. There are multiple functions in a CCA program, each for a different purpose and called with possibly different frequencies. For example, when controlling throughput, the functions to be refined should impact throughput and be called frequently for each acknowledgment packet. 4) The LLM should adjust the correct CCA variables. For example, some Linux CCAs control their throughput using only congestion window size (snd\_cwnd), whereas some others use both congestion window size and the pacing rate (sk\_pacing\_rate). 5) The LLM should determine the correct variable units. The variables in a CCA program have potentially different units and may also change in different versions of the same CCA. 
For example, the Round-Trip Time (RTT) related variables may have units of milliseconds, microseconds, shifted milliseconds, or shifted microseconds.

\subsection{CCA Design Information}\label{sec:input-information}

The CCA design information contains specific user requirements and additional safe network deployment requirements. 

\subsubsection{R1: Minimum throughput requirement}
is a specific user requirement that specifies the minimum throughput that the customized CCA should achieve. It is automatically calculated by the modeling LLM according to the streaming requirement of a user. For example, TABLE~\ref{tab:req-modeling} shows some possible inputs and outputs of the modeling LLM. The first input column shows some possible streaming requirements of non-expert users, and the first output column shows the corresponding minimum bandwidth calculated by the modeling LLM.

\subsubsection{R2: Maximum throughput requirement}
is a requirement for safe network deployment, which limits the maximum throughput of the customized CCA. This ensures that the streaming flow does not starve other flows sharing the home network. It is calculated as a percentage of the Internet connection speed of the home network. For example, if the percentage is 50\%, the streaming flow can use up to 50\% of the Internet connection. The second input column of TABLE~\ref{tab:req-modeling} shows some possible Internet connection speeds, and the second output column shows the corresponding maximum throughput.

\begin{table}[ht]
    \centering
    \renewcommand{\arraystretch}{1.2} 
    \resizebox{\linewidth}{!}{%
    \begin{tabular}{|>{\centering\arraybackslash}p{5cm}|>{\centering\arraybackslash}p{1.5cm}|>{\centering\arraybackslash}p{1.5cm}|>{\centering\arraybackslash}p{2cm}|}
        \hline
        \multicolumn{2}{|c|}{INPUT} & \multicolumn{2}{c|}{OUTPUT} \\
        \hline
        User’s streaming requirement & Home upload speed & Req. throughput &  Max throughput limit \\
        \Xhline{1px}
        \makecell{``HD streaming''} & \makecell{30 Mbps} & \makecell{5 Mbps} & \makecell{15 Mbps} \\
        \hline
        \makecell{``1080p resolution streaming using h.265''} & \makecell{50 Mbps} & \makecell{8 Mbps} & \makecell{25 Mbps} \\
        \hline
        \makecell{``2K resolution 60fps streaming''} & \makecell{80 Mbps} & \makecell{16 Mbps} & \makecell{40 Mbps} \\
        \hline
        \makecell{``4K resolution 60Hz at 30Mbps bitrate''} & \makecell{100 Mbps} & \makecell{30 Mbps} & \makecell{50 Mbps} \\
        \hline
    \end{tabular}
    }
    \caption{Possible inputs and outputs of the modeling LLM}
    \label{tab:req-modeling}
\end{table}

\subsubsection{R3: Mandatory throughput reduction with persistent loss}

is an additional requirement for safe network deployment, which overrides requirement R1 to avoid congestion collapse in the case of persistent packet loss. Specifically, if the cumulative packet loss is higher than a threshold (say 5\% in our experiments), the customized CCA should reduce its throughput as the original CCA.

\subsection{Additional Information}\label{sec:input-additional-information}

A prompt contains additional information for setting LLM parameters, such as the model and temperature.
A prompt also contains additional information about the existing CCA code to refine and relate code references. The code references are essential for code refinement and generation~\cite{2024duevaluating}, including the definition and declaration of functions, structures, and variables. 
To choose the sufficient and non-redundant reference definitions for CCA generation, we choose to extract a subset of the Linux network stack as our reference, which is inspired by \cite{Wei_Sym_ICC_2018} and the legacy Linux Kernel BPF Cubic implementation.

\section{Feedback to Implementation LLM} \label{sec:feedback-iterative-revision}

To fix the possible erroneous output of the implementation LLM, NECC
iteratively asks the LLM to refine the CCA code based on the
network domain-specific feedback. 

\subsection{Existing General Feedback}

Reinforcement through iterative feedback 
\cite{NEURIPS2023_7cc1005e} has been used to improve output quality. Several feedback construction methods have been used in previous works, such as performance assertion~\cite{NEURIPS2023_91edff07}, problem detection request~\cite{NEURIPS2023_7cc1005e}, failed unit test result~\cite{NEURIPS2023_1b44b878}, and static analysis and runtime results~\cite{liu_refining_2024}.

\subsection{Networking Domain-Specific Feedback}\label{sec:feedback-solution}

We have identified three types of domain-specific feedback that directly determine the code quality of generated CCA programs. We also define the satisfaction score between 0\% and 100\% to measure the code quality of a CCA program. A satisfaction score of 100\% means that the NECC generates a functional CCA program that successfully compiles, passes the BPF check, and meets all the user and network safety requirements in all network experiments.

\textbf{F1: Compilation Error Feedback} contains the compilation errors reported by the CCA compiler. The satisfaction score of a program is 0\% if it fails to compile; is 20\% otherwise.

\textbf{F2: BPF Error Feedback} contains the BPF errors reported by the BPF static analysis tool bpf-tool, which ensures that the program is safe for the Linux kernel. The satisfaction score of a program is 40\% if it compiles and passes the BPF check.

\textbf{CCA Performance Feedback} contains the feedback on the CCA performance obtained using the network emulator Mininet. 
NECC runs three groups of network emulation experiments to check the three requirements under short and long RTT network conditions.
\textbf{F3: Feedback on requirement R1.} NECC emulates various congested networks to check if the customized CCA program meets requirement R1; 
\textbf{F4: Feedback on requirement R2.} NECC emulates various non-congested networks to check if the customized CCA program meets requirement R2; 
\textbf{F5: Feedback on requirement R3.} NECC emulates various networks with persistent packet loss to check if the customized CCA program meets requirement R3. The satisfaction score of a program increases by 20\% for each passed group of experiments.

\section{Implementation}\label{sec:Implementation}

We use Python to implement the proposed NECC framework. The specific requirements and home network information of a user are collected using Dify.ai. 
The implementation LLMs are called via their APIs in Python. 
Additionally, we choose Mininet for network emulation to check the user and network safety requirements.

We maintain a group of existing Linux CCA programs, including Reno, Cubic, Vegas, and Illinois, which are used in our experiments. Currently, this group does not include BBR, because there is currently no Linux BPF BBR program due to the complicated BBR functions.

The current NECC implementation makes the following assumptions so that we can focus more on the framework design instead of the framework development details: 1) We assume that the bottleneck of a home network is the Internet connection. 2) We assume that the streaming application of a user runs on a computer with a customized CCA, while all other computers and network devices use their original CCAs.

Potential overhead: According to the statements and experimental results from the Linux BPF developers, the performance of using the BPF-format Cubic is not significantly different from that of the kernel Cubic implementation \cite{bpf}. Therefore, we believe that leveraging the BPF interface does not introduce notable overhead. 

The source code and prompts of NECC and all experiments are published on GitHub at https://github.com/zmrui/NECC.

\section{Evaluation}\label{sec:evaluation}

In this section, we run experiments to study the following research questions about NECC:

\begin{itemize}
    \item \textbf{\textit{RQ1:}} What is the impact of LLM parameters?
    \item \textbf{\textit{RQ2:}} How effective is CoT prompting compared to 0-shot prompting?
    \item \textbf{\textit{RQ3:}} What is the impact of the pool size?
    \item \textbf{\textit{RQ4:}} How effective is the feedback?
    \item \textbf{\textit{RQ5:}} Do customized CCA programs improve live streaming quality?
\end{itemize}

We have evaluated multiple streaming requirements and multiple home Internet speeds and observed similar conclusions. Due to the page limit, we present only the experimental results of 2K at 60 fps streaming and a home Internet speed of 60 Mbps, unless otherwise noted.

\subsection{RQ1: Impact of LLM parameters} \label{sec:eva-llmparam}

\begin{table}[htb]
    \centering
    \renewcommand{\arraystretch}{1.5} 
     \resizebox{\linewidth}{!}{%
     \large
    \begin{tabular}{|p{2cm}|p{2cm}|p{2cm}|p{2cm}|p{2.5cm}|p{2cm}|p{2cm}|}
        \hline
        \makecell{Parameters} & \makecell{Modify \\ inappropr. \\ function} & \makecell{Change \\ only local \\ variables} & \makecell{Only \\ adjust \\ pacing rate} & \makecell{Incorrect \\ throughput \\ formula} & \makecell{Incorrect \\ RTT unit} & \makecell{Incorrect \\ loss rate \\ formula} \\
        \hline
        \makecell{T=0} & \makecell{30} & \makecell{30} & \makecell{0} & \makecell{24} & \makecell{19} & \makecell{0} \\
        \hline
        \makecell{T=0.5} & \makecell{22} & \makecell{4} & \makecell{0} & \makecell{27} & \makecell{6} & \makecell{0} \\
        \hline
        \makecell{T=1} & \makecell{25} & \makecell{6} & \makecell{4} & \makecell{25} & \makecell{8} & \makecell{12} \\
        \hline
    \end{tabular}
    }
    \caption{Design Quality of Cubic. Temperature T=0.5 results have fewer design choice mistakes than T=0 and 1.}
    \label{tab:designquality}
\end{table}

\begin{figure}[tb]
    \centering
    \includegraphics[width=0.9\linewidth]{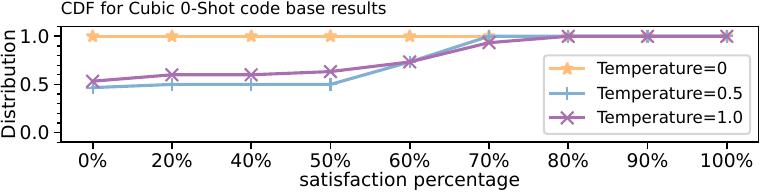}
    \includegraphics[width=0.9\linewidth]{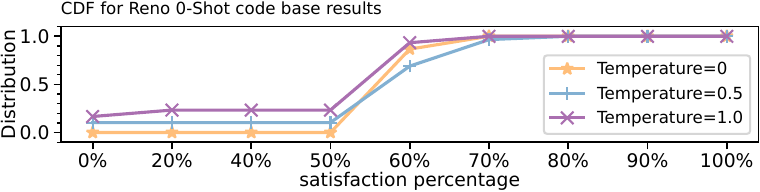}
    \includegraphics[width=0.48\linewidth]{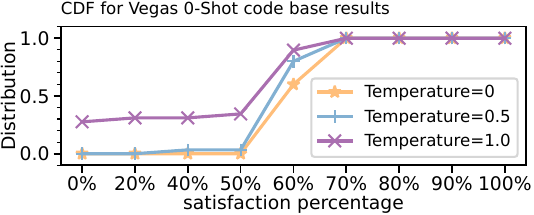}
    \includegraphics[width=0.48\linewidth]{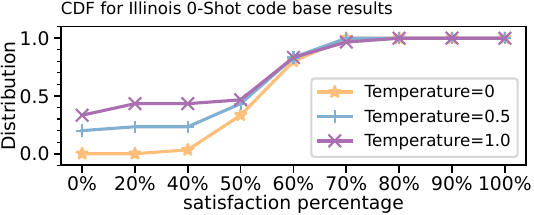}
    \caption{Code quality CDF in different temperatures: temperature 0.5 performs better than temperature 1.}
    \label{fig:rq1parameters}
\end{figure}

We study the impact of two     LLM parameters: model and temperature by using three popular LLM models: GPT-4o (gpt-4o-2024-08-06), GPT-4o-mini (gpt-4o-mini-2024-07-18), and Claude-3.5-sonnet (claude-3-5-sonnet-20240620), and adjusting their temperature settings from 0.0 to 0.5, and 1.0.
For each combination of model and temperature, we use the prompt described in Section~\ref{sec:prompt} to generate a pool of 30 customized CCA programs using Cubic, Reno, Vegas, and Illinois as existing CCA.  Below we present only the GPT-4o results, because Claude-3.5-sonnet sometimes does not follow our instruction, i.e. output the full source code after modification, and GPT-4o always performs better than GPT-4o-mini.

\subsubsection{Design quality}
We manually assess the \textbf{design quality} of a generated CCA program and summarize our results in TABLE~\ref{tab:designquality}, where each row is based on 30 Cubic code candidates.
``Modify inappropriate function'' indicates that the LLM modifies an inappropriate function. For example, it is more effective to modify cubic\_cong\_avoid() than bictcp\_update(), because cubic\_cong\_avoid() is called per ACK, whereas bictcp\_update() is executed only intermittently. ``Change only local variables'' indicates that the LLM adjusts only function-local variables that do not impact TCP throughput, rather than TCP socket variables.
``Adjust only pacing rate''  indicates the LLM adjusts only sk\_pacing\_rate but not snd\_cwnd. This modification is unsuitable for Cubic, as Cubic controls its throughput primarily using snd\_cwnd.
``Incorrect throughput formula'' indicates that the LLM uses an incorrect formula to calculate TCP throughput. For example, the formula does not take the maximum segment size into consideration.
``Incorrect RTT unit'' indicates that the LLM uses an incorrect unit for an RTT variable. For instance, the unit of  srtt\_us is actually 1/8 microsecond instead of 1 microsecond.
``Incorrect loss rate formula'' indicates that the LLM miscalculates the loss rate. For example, the loss rate should be computed as the ratio of lost packets to the total number of transmitted packets rather than to the number of successfully transmitted packets.

\textit{We observe that non-zero temperatures are more likely to generate CCA programs with more diverse designs than a zero temperature for all LLM models}.
Specifically, an LLM model with a zero temperature generates a pool of similar CCA programs, all of which have poor design quality as they modify the same incorrect functions,  adjust the same incorrect CCA variables, and/or use the same incorrect variable units. 
In contrast, an LLM model with a non-zero temperature generates a pool of different CCA programs, some of which have good design quality. 
\textit{Furthermore}, the LLM model with a high temperature (T=1) generates more design mistakes, such as using erroneous formulas or non-existent variables.
Therefore, we choose to select non-zero temperature parameters and prefer parameters that lead to fewer design choice mistakes.

\subsubsection{Code quality}
We use automatically measured satisfaction scores as \textbf{code quality}, which measure the requirement satisfaction degree (e.g., compilation, BPF, and performance) of NECC-generated CCA programs as described in Section~\ref{sec:feedback-iterative-revision}.
Fig. \ref{fig:rq1parameters} shows code quality results through the Cumulative Distribution Function (CDF) of the satisfaction scores, where the x-axis is the satisfaction score, and the y-axis is the CDF $F(x)=P(X\leq x)$. The CDF curves approaching the bottom right are more desirable, indicating that more programs achieve higher satisfaction scores and thus better code quality. 
\textit{We observe that temperature 0.5 code quality performs better than temperature 1}, which is the same as the design quality result preference.

\subsubsection{Discussion}
We choose the non-zero, intermediate value of 0.5 as the temperature parameter based on the design quality and code quality evaluation results.

\subsection{RQ2: Effectiveness of CoT Prompting}\label{sec:RQ2Cot}

\begin{figure}[tb]
    \centering
    \includegraphics[width=0.48\linewidth]{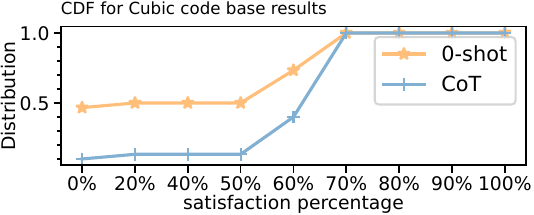}
    \includegraphics[width=0.48\linewidth]{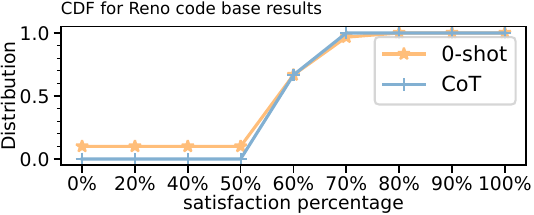}
    \includegraphics[width=0.48\linewidth]{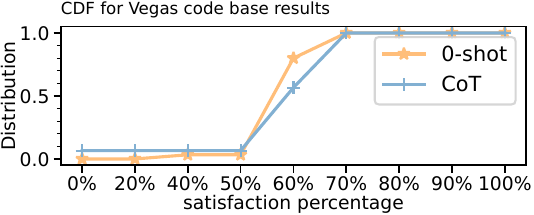}
    \includegraphics[width=0.48\linewidth]{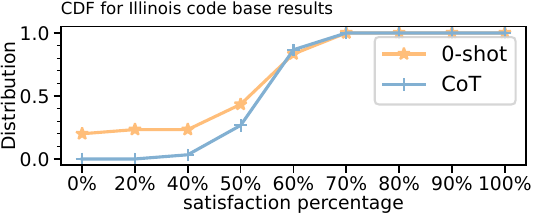}
    \caption{CoT may or may not improve code quality compared with 0-shot prompting, mainly due to  compilation errors.}
    \label{fig:Cotoshot}
\end{figure}

We evaluate the effectiveness of CoT prompting by comparing it with 0-shot prompting. 
We use CoT or 0-shot prompting to generate a pool of 30 customized CCA programs for each of the following CCAs: Reno, Cubic, Vegas, and Illinois. 
We measure both design quality and code quality. 

\textit{We observe that CoT prompting effectively improves the design quality of the generated CCA program compared with 0-shot prompting.}
With the guidelines contained in the CoT promotes described in Section~\ref{sec:prompt}, the LLM is more likely to select the appropriate CCA functions to modify, adjust the correct CCA variables, and use the proper variable units.

\textit{We  observe that CoT prompting may or may not improve code quality compared with 0-shot prompting, mainly due to potential compilation errors.} 
Fig. \ref{fig:Cotoshot} shows CDF of the satisfaction scores of a pool of CCA programs using 0-shot or CoT prompting.
CoT prompting improves the code quality for Cubic and Illinois, maintains the same code quality for Reno, and degrades the code quality for Vegas compared with 0-shot prompting. 
We find that the variability in code quality with CoT prompting is primarily due to potential compilation errors. Although our CoT prompts instruct the LLM to generate CCA programs without compilation errors, it may still occasionally generate CCA programs that fail to compile. 
Even a single compilation error results in a zero satisfaction score, regardless of the design quality.

\begin{figure}[t]
    \centering
    \includegraphics[width=0.9\linewidth]{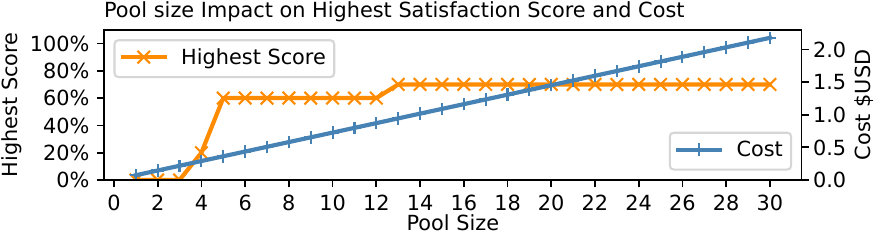}
    \caption{A larger pool size improves the highest satisfaction score of the pool, but it also leads to a higher cost.}
    \label{fig:poolsize}
\end{figure}

\subsection{RQ3: Impact of Pool Size}

We use this group of experiments to study the impact of pool size on code quality and cost. We generate 30 candidates and sort their satisfaction scores incrementally, representing $n \in [1, 30]$ customized CCA programs. 
We then measure the highest satisfaction score and assess the cost in U.S. dollars for each pool size \texttt{n}.  

Fig.~\ref{fig:poolsize} shows the results for Cubic using CoT prompting.  
\textit{We observe that a larger pool size improves the highest satisfaction score of the pool, but it also leads to a higher cost.}
This is because a larger pool size yields more diverse candidates and thus increases the likelihood of good candidates.
However, the cost of pre-trained LLMs must be taken into account. 
For instance, an initial request containing the Cubic source code and relevant references without any conversation history 
costs \$0.018, while a typical customized CCA program costs \$0.072 on gpt-4o-2024-08-06; the sending and receiving call process of pre-trained LLM via Python API costs about 40 seconds, and the emulation will cost about 5 minutes.
In other experiments, we use a pool size of 5 because this is the pool size that balances the highest satisfaction score with the cost. 
We also observe similar trends for other CCAs, such as Reno, Vegas, and Illinois. However, even with a large pool size, CCAs may not achieve the 100\% highest satisfaction score.

\subsection{RQ4: Effectiveness of Feedback}\label{sec:eva-feedback}

\begin{figure}[t]
    \centering
    \includegraphics[width=0.9\linewidth]{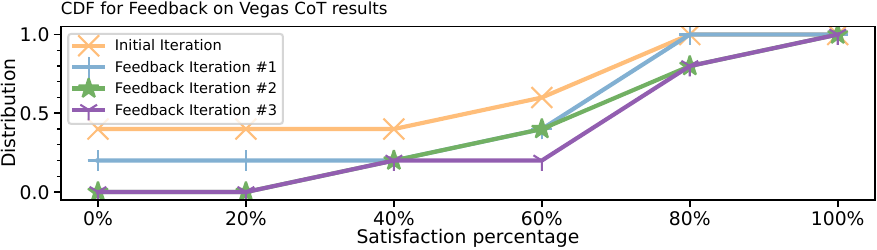}
    \caption{Feedback effectively improves the code quality over iterations. }
    \label{fig:feedbackvegas}
\end{figure}

We evaluate the effectiveness of feedback by analyzing its impact on code quality at each iteration.
Fig.~\ref{fig:feedbackvegas} shows the satisfaction score CDF of a pool of CCA programs for Vegas at multiple iterations. The pool in the initial iteration is generated using CoT promoting.  We observe similar trends for other CCAs, though the extent of improvement varies.

\textit{We observe that proposed feedback in Section~\ref{sec:feedback-iterative-revision} effectively improves the code quality progressively over the iterations. } For example, some programs have a score of zero due to compilation errors in the initial iteration, and all programs successfully compile without compilation errors after two iterations. Additionally, the highest score of all programs in the pool is 80\% in the initial iteration, and it reaches 100\% after two iterations. By comparing different types of feedback, we find that the compilation feedback (i.e., F1) and BPF feedback (i.e., F2) are more effective than CCA performance feedback (i.e., F3, F4, and F5). For example, all programs in the pool successfully pass both the compilation and BPF checks after two iterations, but some programs still fail one or more CCA performance checks. This is because compilation and BPF feedback are more specific (e.g., errors for specific functions or variables) than CCA performance feedback. 

\subsection{RQ5: Impact on streaming experience}

Finally, we study the impact of a customized CCA program on the streaming experience. We use FFmpeg to emulate the live streaming of a game with an average bitrate of 30 Mbps. The home has an Internet speed of 60 Mbps and an RTT of 50 ms, which is shared by the streaming flow and $a$ additional TCP flows. The value of $a$ varies from 0 to 4 to emulate different levels of congestion ($a \leq 1$: not congested, $a=2$: lightly congested, $a \geq 3$: highly congested). The streaming video is received and played using a VLC player, and the streaming quality is measured using the Structural Similarity Index Measure (SSIM).  For live streaming users, SSIM $\in [0.9, 1.0]$ is usually considered good, $[0.8, 0.9]$ is fair, and $[0.0, 0.8]$ is unsatisfactory. 

\begin{figure}[t]
    \centering
    \includegraphics[width=0.9\linewidth]{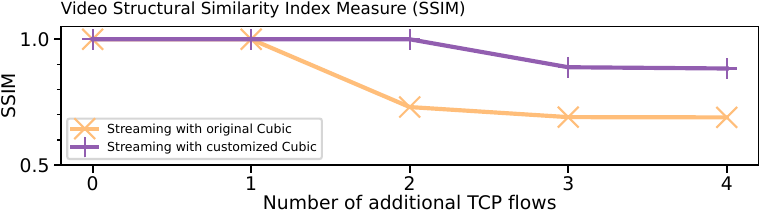}
    \caption{Customized Cubic achieves significantly better SSIM for live streaming than the original Cubic in both lightly and highly congested networks.}
    \label{fig:ssim}
\end{figure}

Fig.~\ref{fig:ssim} shows the SSIM values of live streaming using a customized Cubic generated by NECC or using the original Linux Cubic. We observe that \textit{the customized Cubic achieves significantly better SSIM than the original Cubic in congested networks.} The customized Cubic achieves good SSIM in both lightly and highly congested networks, whereas the original Cubic achieves unsatisfactory SSIM in all congested networks.

\subsection{Insights, Discussions, and Limitations}

Our experiments show that CoT prompting is complementary to feedback because CoT prompting with general guidelines is more effective in improving high-level design quality, whereas feedback with detailed errors is more effective in improving the actual code quality.

We also notice that there are some inherent reasons for the erroneous outputs of LLMs. 1) \textit{Incomplete information}: Limited by the size of a prompt and the extremely long source code of the Linux kernel TCP, a prompt can contain only part of Linux TCP code and thus the LLMs may misunderstand how the CCA works (e.g., mistakenly adjusting only the pacing rate \texttt{sk\_packing\_rate} but not the congestion window size \texttt{snd\_cwnd}). The incomplete information can be potentially addressed by providing the LLMs with more domain knowledge in a prompt. 2) \textit{Ambiguous information}: Existing CCA code may have some comments that are ambiguous to the LLMs (e.g., $<<$ in comments of C programs may be used to highlight text or used as a left shift operation). The ambiguous information can potentially be addressed by generating a pool of candidate CCA programs and setting the LLM temperature to a non-zero value, as NECC does. By doing so, NECC can explore different understandings. 3) \textit{Lack of performance input and output examples}: The LLMs may or may not correctly refine CCA programs based on the performance feedback, because they do not know exactly how and how much each line of the code impacts the final performance. This can potentially be addressed by providing the LLMs with a comprehensive set of performance input and output examples, in addition to performance feedback.

Current NECC considers only two network safety requirements and makes some assumptions about home networks. We plan to consider more network safety requirements and eliminate these assumptions so that the generated customized CCA code can be widely and safely deployed on the Internet. Current NECC considers only throughput requirements of live-streaming users. In the future, we plan to explore other types of user requirements, such as latency requirements, and other types of users, such as gaming users. Current NECC considers only four existing CCAs for Linux BPF. In the future, we plan to extend NECC to more existing CCAs, such as BBR, and other user-space CCAs, such as those with QUIC.

\section{Conclusion}

We presented a non-expert customized CCA framework called NECC, which enables non-expert users to model,
implement, and deploy their customized CCAs.
Our evaluations using real-world CCAs show that the performance of NECC is very promising.

\section{Acknowledgment}

This work was supported in part by NSF CCF-2124116 and CNS-2135539.

\bibliographystyle{IEEEtran}
\bibliography{refs/AI,refs/SE,refs/networking,refs/misc}

\end{document}